\documentclass[twocolumn,showpacs,preprintnumbers,floatfix,amsmath,amssymb,
superscriptaddress]{revtex4}

\usepackage{graphicx}
\usepackage{dcolumn}
\usepackage{bm}

\newcommand{\be}{\begin{equation}}
\newcommand{\ee}{\end{equation}}
\newcommand{\bn}{\begin{eqnarray}}
\newcommand{\en}{\end{eqnarray}}
\newcommand{\ba}{\begin{array}}
\newcommand{\ea}{\end{array}}
\newcommand{\bc}{\begin{center}}
\newcommand{\ec}{\end{center}}
\newcommand{\bml}{\begin{mathletters}}
\newcommand{\eml}{\end{mathletters}}

\begin{document}

\preprint{ }

\title{Shell Model and Mean-Field  Description of Band Termination
}

\author{M. Zalewski}
\affiliation{
Institute of Theoretical Physics, University of Warsaw, ul. Ho\.za 69,
00-681 Warsaw, Poland }%

\author{W. Satu{\l}a}
\affiliation{
Institute of Theoretical Physics, University of Warsaw, ul. Ho\.za 69,
00-681 Warsaw, Poland }%
\affiliation{Joint Institute for Heavy Ion Research,
Oak Ridge National Laboratory,  Oak Ridge, Tennessee 37831}

\author{W. Nazarewicz}
\affiliation{
Department of Physics and Astronomy, University of Tennessee, Knoxville, Tennessee
37996
}%
\affiliation{
Physics Division, Oak Ridge National Laboratory, Oak Ridge, Tennessee 37831
}%
\affiliation{
Institute of Theoretical Physics, University of Warsaw, ul. Ho\.za 69,
00-681 Warsaw, Poland }%

\author{G. Stoitcheva}
\affiliation{Lawrence Livermore National Laboratory, P.O. Box 808, L-414, Livermore, California 94551
}%

\author{H. Zdu\'nczuk}
\affiliation{
Institute of Theoretical Physics, University of Warsaw, ul. Ho\.za 69,
00-681 Warsaw, Poland }%

\date{\today}

\begin{abstract}
We study  nuclear high-spin states undergoing the
transition to the fully stretched  configuration  with
maximum angular momentum $I_{max}$ within the space of valence nucleons.
To this end, we perform a systematic theoretical analysis  of non-fully-stretched
$I_{max}-2$ and $I_{max}-1$  $f_{7/2}^n$ seniority isomers
and $d_{3/2}^{-1} f_{7/2}^{n+1}$ intruder
states in the $A$$\sim$44 nuclei from the lower-$fp$ shell.
We employ two theoretical approaches:
(i) the density functional theory  based on the cranked
self-consistent Skyrme-Hartree-Fock method, and (ii)
the nuclear shell model
in the full  $sdfp$ configuration space allowing for 1p-1h
cross-shell excitations.
 We emphasize the importance of
restoration of  broken  angular momentum
symmetry inherently obscuring
the mean-field   treatment of high-spin states.
Overall good agreement with
experimental data is obtained.
\end{abstract}


\maketitle

\section{Introduction}
The phenomenon of band termination is a splendid  manifestation of a
competition between nuclear single-particle and collective motion. At
low angular momenta, a rotational band is associated with a collective
reorientation of a deformed nucleus in space, with many nucleons
contributing coherently to the total spin. With increased rotational
velocity, however, the
Coriolis interaction causes  nucleonic pairs to break, and the decoupled
nucleons align their  individual angular momenta. It often
happens that breaking relatively few nucleonic pairs  can give rise to a
 nuclear state with a fairly large spin. In the language of
the nuclear shell model, of particular importance are the ``seniority
isomers" or ``fully-aligned (or stretched, or optimal) states," which carry the
maximum angular momentum $I_{max}$ within the  space of valence nucleons.

The transition process from collective rotation at high spins to a
single-particle picture at $I_{max}$  is referred to as band termination
 \cite{[Boh75]}. Terminating bands are common in nuclei; they have been
observed across the table of the nuclides (see recent reviews
\cite{[Afa99a],[Sat05]}). The nature of the
termination process strongly depends on the size of the valence space. For instance,
if only several valence nucleons are present, the static nuclear
deformation cannot develop and the collective effects
have dynamic character. In this case, $I_{max}$ can
be generated by breaking very few nucleonic pairs, and the transition to
the single-particle limit is rapid. On the other hand, for deformed nuclei having many
valence nucleons,  the transition to the non-collective regime is long and
gradual, often involving many intermediate stages.

Theoretically, high-spin band terminations and fully aligned
configurations are often discussed within the nuclear
shell model (SM) and/or the self-consistent density functional theory (DFT).
The nuclear shell model \cite{[Cau05]} can be applied
to nuclei with several valence nucleons outside the magic core.
The effective Hamiltonian is exactly diagonalized in a
subspace of  many-body Slater determinants  resulting in
correlated wave functions that preserve angular momentum, parity,
 particle number, and - usually - isospin.

For medium-mass and heavy nuclei containing
 many valence nucleons,   the tool of choice is
the nuclear DFT~\cite{[Ben03]}.
Here,  the nucleus is described in terms of
 one-body densities and currents representing
distributions of nucleonic matter, spins, momentum, and kinetic energy,
as well as their derivatives. The associated mean fields,
obtained by means of the self-consistent
Hartree-Fock (HF) procedure, are usually deformed,
i.e., they spontaneously break the symmetries
of the underlying  Hamiltonian. In this way, many
essential many-body correlations can be incorporated into
a single product state \cite{[Sat05],[Ben03],[RS80]}.
 However, the price paid for the simple intrinsic picture is high:
the HF wave function is no longer an eigenstate of symmetry
operators; hence, the transformation to the laboratory reference
frame has to be carried out to restore broken symmetries.

The language in which the nucleus is pictured as a wave packet with
anisotropic density/current distribution is particularly useful for the
description of rotational motion within the cranking formalism. Here,
the many-body Hamiltonian $\hat{H}$ is replaced by a  rotating
Hamiltonian (Routhian), $\hat H^\omega = \hat H - \omega \hat J_z$, where the
rotational frequency $\omega$ is interpreted as a Lagrange multiplier
determined from the angular momentum constraint. The resulting cranked
HF method (CHF) can  also be used to describe the so-called {\it
non-collective\/} rotations, i.e., rotations around symmetry
axis~\cite{[Afa99a],[Szy83]}. The
non-collective CHF technique applied in this work has proven to be a
very efficient and accurate way to generate stretched solutions within a
self-consistent framework.

In practice, most applications of nuclear  DFT are based on  Skyrme
energy density functionals optimized to various experimental data
\cite{[Ben03],[Sto06a]}. Recently, such an approach
was applied~\cite{[Zdu05x],[Sto06]} to fully-stretched,
 high-spin  states associated with the $[f_{7/2}^n]_{I_{max}}$ and
$[d_{3/2}^{-1} f_{7/2}^{n+1}]_{I_{max}}$  SM configurations ($n$ denotes
the number of valence particles outside the $^{40}$Ca core) of $20\leq Z
\leq N\leq 24$. These studies have demonstrated that those fully-aligned
states have a fairly simple SM structure, and, therefore,  they provide an
excellent  testing ground for both the time-odd densities and fields that
appear in the mean-field  description and for SM effective
interactions. In particular, it was shown  that the energy difference
between the excitation energies of the terminating states,
$E([d_{3/2}^{-1} f_{7/2}^{n+1}]_{I_{max}}) - E([f_{7/2}^n]_{I_{max}})$,
is a sensitive probe of time-odd spin couplings and the strength of the
spin-orbit term in the  Skyrme functional
\cite{[Zdu05x]}, and that with properly modifying
functionals, the nuclear DFT provides  a description of   the data for
stretched states that is of similar quality as the fully correlated
SM~\cite{[Sto06]}.

The aim of this study is to understand the nature of
the $[f_{7/2}^{n}]_{I_{max}-1}$  (Sec.~\ref{Im1a}) and
$[d_{3/2}^{-1} f_{7/2}^{n+1}]_{I_{max}-1}$ (Sec.~\ref{Im1b})
configurations in the $A$$\sim$44 mass region.
These $I_{max}-1$ states, usually referred to  as
unfavored-signature terminating states, can be obtained
from  fully-stretched
configurations by signature-changing particle-hole (p-h)
excitations. Consequently, their structure is sensitive
to both shell structure and time-odd nuclear fields. We
demonstrate that  non-collective
cranking may lead to a dramatic violation of rotational
symmetry even for the cases when the nuclear shape  is almost spherical.
We also analyze in Sec.~\ref{Im2}  the $I_{max}-2$ weakly collective states in
 terminating normal-parity and intruder structures. The
 associated correlation energy, mainly  associated with quadrupole
 effects, is calculated to be large, around 2\,MeV.

\section{The models}\label{models}
The details of DFT and SM frameworks applied in this study
follow  Ref.~\cite{[Sto06]} in which
the stretched configurations in the $A$$\sim$44
mass region have been successfully explained.
The CHF calculations are carried out using
the HFODD code of Ref.~\cite{[Dob04t]}. We employed the
SLy4~\cite{[Cha97]} and SkO~\cite{[Rei99]}  Skyrme energy density
functionals  slightly modified along the prescription of
Refs.~\cite{[Zdu05x]}. Without going into detail, the
modifications concern the coupling constants of ${\boldsymbol s}^2$ and
${\boldsymbol s}$$\cdot$$\Delta{\boldsymbol s}$ terms giving rise to
the time-odd spin mean fields. This has been done by
constraining the functionals to the empirical
spin-isospin Landau  parameters.
 In addition, the strength of the
spin-orbit interaction has been reduced by 5\% from the original SLy4 and
SkO values. As discussed
 in Ref.~\cite{[Sto06]}, the DFT results
for the $T$=0 states in $N$=$Z$ nuclei have to be corrected for
the isospin breaking effects. Here we assume that the isospin  correction
weakly depends on angular momentum; hence, it has been neglected when discussing
energy differences between high-spin states.

During the last decade, extensive SM calculations for the light
$fp$-shell nuclei
have been conducted using different SM interactions
\cite{[Bed97],[Pov98],[Bra00],[Has00],[Jua06]}. In most cases, SM
reproduces well the excitation energies of normal-parity and
intruder states, as well as transition
rates. In general, the intruder $d_{3/2}^{-1} f_{7/2}^{n+1}$ states are
predicted and observed to be more collective than the
$f_{7/2}^n$ structures, i.e.,
the associated in-band $E2$ rates are significantly greater within
intruder bands.
The SM results in this work are obtained using the code
ANTOINE~\cite{[Cau95]} in the $sdfp$ configuration space limited to
1p-1h cross-shell excitation from the $sd$ shell to the $fp$ shell. In
the $fp$-shell SM space we took  the FPD6 interaction \cite{[Ric90]}.
The remaining matrix elements are those of Ref.~\cite{[War90]}. As
compared to the earlier work~\cite{[Bed97]}, the mass scaling of the SM
matrix elements was done here consistently, thus reducing  the $sd$
interaction channel by $\sim$4\%.

\section{Unfavored-signature terminating $f_{7/2}^{n}$ states}\label{Im1a}
\begin{figure}[htb]
\centerline{\includegraphics[
width=0.45\textwidth,clip]{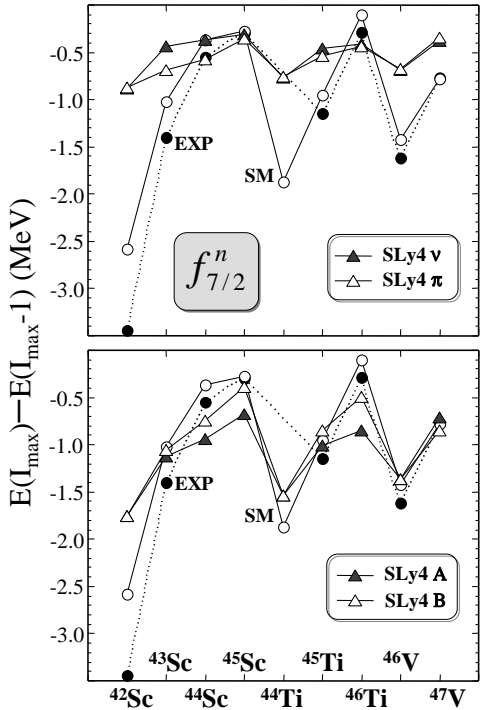}}
\caption{The  energy difference
between $I_{max}$ and $I_{max}-1$ states in  $f_{7/2}^{n}$ configurations in
$A$$\sim$44 nuclei. Dots and circles mark experimental data and the SM
results, respectively. Top: CHF-SLy4 results for unfavored signature
terminating states corresponding to neutron (filled triangles) and proton
(open triangles) signature inversion. Bottom: CHF-SLy4 results including the
angular momentum correction calculated using prescription
A (filled triangles) and B (open triangles).}\label{f72}
\end{figure}

We begin with
$[f_{7/2}^{n}]_{I_{max}-1}$ configurations
in  $20 < Z \leq N \le 24$ nuclei.
Figure~\ref{f72}  displays the energy difference between the stretched
 $I$=$I_{max}$ and the lowest  $I_{max}-1$ states.
It is gratifying to see
a qualitative, and -- in most cases --  quantitative
agreement between SM and experimental data.

Within the {\it na\"{\i}ve} non-collective cranking, the
unfavored $I_{max}-1$
states can be obtained by either  inverting the signature of a single proton
($\pi$) or a single neutron ($\nu$). The energies of those states,
$E^{(\textrm{CHF},\pi)}_{I_{max}-1}$ and
$E^{(\textrm{CHF},\nu)}_{I_{max}-1}$, are displayed in Fig.~\ref{f72}
(top) with respect to the energy $E^{(\textrm{CHF})}_{I_{max}}$ of the
stretched configuration. It is seen that agreement between CHF and
experiment is rather poor. In particular, the strong particle number
variation in the energy difference is not reproduced by CHF.
This discrepancy has its origin in the spontaneous violation of
rotational invariance by the mean-field solutions, in spite of the fact
that the underlying CHF states  are almost spherical in all the
considered cases. Indeed, since  the cranking  procedure only constrains an
expectation value of the angular momentum projection,
CHF states contain components with different angular momentum.
In the case of unfavored $I_{max}-1$ states,
 two components are expected to dominate: spurious  reorientation mode $|I_{max};
I_{max}-1\rangle$  and physical stretched  configuration $|I_{max}-1;
I_{max}-1\rangle$.
%

The appearance of spurious  components in partially-aligned cranking
configurations  is well known \cite{[Abe85]}. A classic example is the
cranking treatment of two identical nucleons  in a spherical single-$j$
shell. While the stretched configuration with $J_z$=$M$=$2j-1$ can be
associated with the stretched state having $I_{max}$=$2j-1$, the cranked
$J_z$=$2j-2$ solution is simply  a magnetic substate of $I_{max}$ (the
$I_{max}-1$ state does not exist in a $j^2$ configuration as a result
 of the
Pauli principle). In order to remove spurious components, angular
momentum needs to be restored. Since the number of $M$-scheme
configurations around $I_{max}$ is very limited, in our cranking
analysis we resort to an approximate projection scheme. In
the following, the adopted angular momentum restoration procedure
is discussed for the case of $^{43}$Sc .

Due to a
near-sphericity of the CHF solutions, single-particle (sp)
states can be labeled using the
angular momentum projection $m$.
The optimal state $|I_{max}; I_{max}\rangle$ can be viewed
as a CHF vacuum around which p-h excitations are built.
The $|I_{max}; I_{max}-1\rangle$ spurious state can be obtained by acting on
the vacuum with  the lowering operator $\hat I_{-}$. For $^{43}$Sc, which
in a  SM picture has one proton and two neutrons in a $f_{7/2}$ shell,
$I_{max}^\pi$=19/2$^-$ and the (unnormalized) spurious state can be written as
\begin{eqnarray}\label{spur1}
|I_{max}; I_{max}-1 \rangle   =
         2\sqrt{3}\, \hat a^\dagger_{\nu\, 3/2} \hat a_{\nu\, 5/2}
         |I_{max}; I_{max}\rangle +  \nonumber  \\
        \sqrt{7}\, \hat a^\dagger_{\pi\, 5/2} \hat a_{\pi\, 7/2}
        |I_{max}; I_{max}\rangle
          \equiv  a\, |\nu\rangle + b\, |\pi\rangle \, ,
\end{eqnarray}
where
$\hat a^\dagger_{\tau\, m}$  ($\tau$=$\pi$ or $\nu$) represents a
particle in the $f_{7/2}$ shell
carrying   magnetic quantum number  $m$. By assuming
that the valence coupling scheme (\ref{spur1}) can be extended to the CHF
case,
the $|I_{max}; I_{max}-1 \rangle$ state can be represented
as a unique combination of  CHF solutions
$|\nu\rangle$ and  $|\pi\rangle$ corresponding to the lowest
neutron and proton p-h signature-changing excitations.

The mixing coefficients $a$ and $b$ introduced in  Eq.~(\ref{spur1})
for $^{43}$Sc
can be calculated in a similar  manner for
any nucleus. They are displayed in Table~\ref{wyn}.  Assuming that
contributions from  other shells are small,
the  physical state $|I_{max}-1; I_{max}-1 \rangle$
can be represented  by the orthogonal combination:
\begin{equation}\label{spur2}
|I_{max}-1; I_{max}-1 \rangle  =  -b\, |\nu\rangle + a\, |\pi\rangle\, .
\end{equation}
%
\begin{table}[t]
\begin{center}
\caption[A]{Maximum
spin states $I_{max}^\pi$ and squared
unnormalized expansion amplitudes  for
$f_{7/2}^{n}$
and  $d_{3/2}^{-1}f_{7/2}^{n+1}$  configurations
in $A$$\sim$44 nuclei. In the latter case only $N$$\ne$$Z$ nuclei are
considered.}\label{wyn}
\begin{ruledtabular}
\begin{tabular}{cccccccccc}
        &  $\quad\quad$ & $I_{max}$ & $a^2$ & $b^2$  & $\quad\quad$ &
                                   $I_{max}$ & $a^2$ & $b^2$  & $c^2$
                \\
     $^{42}\textrm{Ca}$    &&           &     &      &  &  $11^-$ &12 &  7  & 3
                \\ 
     $^{44}\textrm{Ca}$    &&           &     &      &  &  $13^-$ &16 &  7  & 3
                \\ 
     $^{42}\textrm{Sc}$    && $7^+$     &  1  &  1   &  &         &   &     &
                \\ 
     $^{43}\textrm{Sc}$    && $19/2^-$  & 12  &  7   &  &$27/2^+$ &12 & 12  & 3
                \\ 
     $^{44}\textrm{Sc}$    && $11^+$    & 15  &  7   &  &  $15^-$ &15 & 12  & 3
                \\ 
     $^{45}\textrm{Sc}$    && $23/2^-$  & 16  &  7   &  &$31/2^+$ &16 & 12  & 3
                \\ 
     $^{44}\textrm{Ti}$    && $12^+$    &  1  &  1   &  &         &   &     &
                \\ 
     $^{45}\textrm{Ti}$    && $27/2^-$  & 15  & 12   &  &$33/2^+$ &15 & 15  & 3
                \\ 
     $^{46}\textrm{Ti}$    && $14^+$    & 16  & 12   &  &  $17^-$ &16 & 15  & 3
                \\ 
     $^{46}\textrm{V}$     && $15^+$    &  1  &  1   &  &         &   &     &
                \\ 
     $^{47}\textrm{V}$     && $31/2^-$  & 16  & 15   &  &$35/2^+$ &16 & 16  & 3
\end{tabular}
\end{ruledtabular}
\end{center}
\end{table}
This $2\times 2$ configuration mixing  can
be dealt with using  two slightly different methods.
In the first variant (called A in the following), one
 requires  that the spurious solution
is degenerate with respect to the CHF optimal state  having energy
$E^{(\textrm{CHF})}_{I_{max}}$. This assumption leads
to a simple expression for the energy of the unfavored-signature
terminating state
\begin{equation}\label{vA}
E^{(\textrm{CHF},A)}_{I_{max}-1}=E^{(\textrm{CHF},\pi)}_{I_{max}-1} +
E^{(\textrm{CHF},\nu)}_{I_{max}-1} - E^{(\textrm{CHF})}_{I_{max}}.
\end{equation}
In the second variant (called B), the
mixing coefficients  are taken directly from Table~\ref{wyn}
and the corresponding energy can be written as
\begin{equation}\label{vB}
E^{(\textrm{CHF},B)}_{I_{max}-1}=
\frac{a^2 E^{(\textrm{CHF},\pi)}_{I_{max}-1} -
b^2 E^{(\textrm{CHF},\nu)}_{I_{max}-1}}{a^2-b^2}.
\end{equation}
Note that this method cannot be applied to $N$=$Z$ nuclei where
$a^2$=$b^2$.  Moreover, method B does not guarantee that the energy of the
spurious
state is degenerate with the optimal state.

The energy differences   between $I_{max}$ and $I_{max}$$-$1 states
obtained in variants A and B are displayed in the lower panel of
Fig.~\ref{f72}. It is seen that the the effect of symmetry restoration
is large. In particular, the  energies corrected for the angular momentum
mixing follow fairly  accurately experiment and SM. Moreover,
methods A and B give fairly similar results, although method B is
slightly closer to the data. Similar results were also obtained in the
CHF-SkO variant.

\section{Unfavored-signature terminating  $d_{3/2}^{-1}f_{7/2}^{n+1}$ states}\label{Im1b}
We now consider  the $[d_{3/2}^{-1}f_{7/2}^{n+1}]_{I_{max}-1}$
intruder configurations. Let us recall that in this case the
$[d_{3/2}^{-1}f_{7/2}^{n+1}]_{I_{max}}$
configurations are uniquely defined only for  $N\ne Z$ nuclei.
Indeed, as discussed in Ref.~\cite{[Sto06]}, the CHF solutions
for $N$=$Z$ systems violate isobaric symmetry and can
no longer  serve  as
reference  states. Hence, we shall limit our
considerations to $20 \le  Z <  N \leq 24$ nuclei. Moreover, in the
following,  we will consider only configurations involving
1p-1h  proton  excitation across the $Z$=20 gap.
The latter assumption concerns the $N$$-$$Z$=1 nuclei in which
the neutron cross-shell excitations can also give
rise to aligned  $I_{max}-1$ states.
It is worth mentioning that the 1p-1h neutron excitations in
$N$$-$$Z$=1 nuclei are higher than the lowest 1p-1h proton excitations.
Moreover, they do not mix at the level of CHF, leading to a severe
isospin symmetry violation. Restoration of the isospin symmetry
for these cases is, however, beyond the scope of this work.

Under these assumptions, within the non-collective cranked CHF picture,
there are three obvious ways of  changing the signature quantum number
of an intruder state.
Namely, one can  invert the signature of a single $f_{7/2}$ proton
($|\pi\rangle$),  a single $f_{7/2}$  neutron ($|\nu\rangle$), or a
proton  $d_{3/2}$ hole ($|\bar{\pi}\rangle$).  In this case, SM yields
two independent $(I_{max}-1)_{i}$ ($i$=1,2)  low-lying solutions. In
CHF, both these solutions are polluted by the presence of spurious
$|I_{max}; I_{max}-1\rangle$ components.

Since the CHF solutions for the aligned intruder states
are nearly  spherical, we can employ
the same technique as proposed earlier  for the $f^n_{7/2}$ stretched states
to approximately restore the
rotational invariance. The only difference is that in addition
to
active $f_{7/2}$ particles,  one has to consider
active $d_{3/2}$ proton holes (denoted as $\bar\pi$ in the following).
As usual, the $|I_{max} , I_{max} -1\rangle$ is obtained
by acting with  $\hat{I}_-$  on the $|I_{max}, I_{max}\rangle$ CHF vacuum.
For the representative case
of $^{42}$Ca  one obtains:
\begin{eqnarray}\label{spur}
   |I_{max} , I_{max} -1 \rangle
        & = & \sqrt{12}\,
      \hat a^\dagger_{\nu\, 3/2} \hat a_{\nu\, 5/2}
                                      |I_{max} , I_{max} \rangle
                \nonumber \\
        & + & \sqrt{7}\,
      \hat a^\dagger_{\pi\, 5/2} \hat a_{\pi\, 7/2}
                                      |I_{max} , I_{max} \rangle \nonumber \\
        & + &   \sqrt{3}\,
      \hat a^\dagger_{\bar\pi\, 1/2} \hat a_{\bar\pi\, 3/2}
                                      |I_{max} , I_{max} \rangle
         \nonumber \\
    &  \equiv &  a|\nu\rangle + b|\pi\rangle + c|\bar\pi\rangle\, .
\end{eqnarray}
Similar calculations can   be carried out  for all $N\ne Z$ nuclei.
The resulting  mixing coefficients are collected in Table~\ref{wyn}.

The intruder case  represents a  $3\times 3$ mixing problem.
Two different analytical approximate  projection techniques
have been  developed \cite{[Zal06]}. In the first method (called A), we assume real
mixing matrix elements and require the eigenvector
$(a,b,c)$ of  Eq.~(\ref{spur}) and Table~\ref{wyn} to correspond
to zero energy mode relative to the CHF energy
$E^{(\textrm{CHF})}_{I_{max}}$ of the optimal
solution. By introducing $e_\alpha= E^{(\textrm{CHF},\alpha)}_{I_{max}-1}-
E^{(\textrm{CHF})}_{I_{max}}$, where $\alpha$=$\nu$ (or 1),
$\pi$ (or 2), and $\hat{\pi}$ (or 3),
the  energies of physical solutions  relative to the CHF optimal state
$E^{(\textrm{CHF})}_{I_{max}}$  are:
\begin{equation}\label{ener}
        \lambda_{\pm} = \frac{1}{2}
        \left(\sum_i e_i \pm \sqrt{ (\sum_i e_i )^2 - 4Z}\right),
\end{equation}
where
\begin{equation}\label{aux}
        Z = \sum_{i<j} (e_ie_j - |V_{ij}|^2)
\end{equation}
and
\begin{eqnarray}\label{int}
        V_{12} &=& \frac{-e_1a^2-e_2b^2+e_3c^2}{2ab}, \nonumber \\
        V_{13} &=& \frac{-e_1a^2+e_2b^2-e_3c^2}{2ac},  \\
        V_{23} &=& \frac{e_1a^2-e_2b^2-e_3c^2}{2bc}. \nonumber
\end{eqnarray}

In the second method (called B), we admit complex mixing amplitudes. In
this case we set the
spurious mode to zero  but do not require the
corresponding eigenvector to be equal $(a,b,c)$. This
procedure leads to exactly the same set of formulas (\ref{ener})-(\ref{aux})
as above  but with $V_{\alpha\beta}=\sqrt{e_\alpha e_\beta}$ (see
Ref.~\cite{[Zal06]} for further details).

\begin{figure}[htb]
\centerline{\includegraphics[
width=0.45\textwidth,clip]{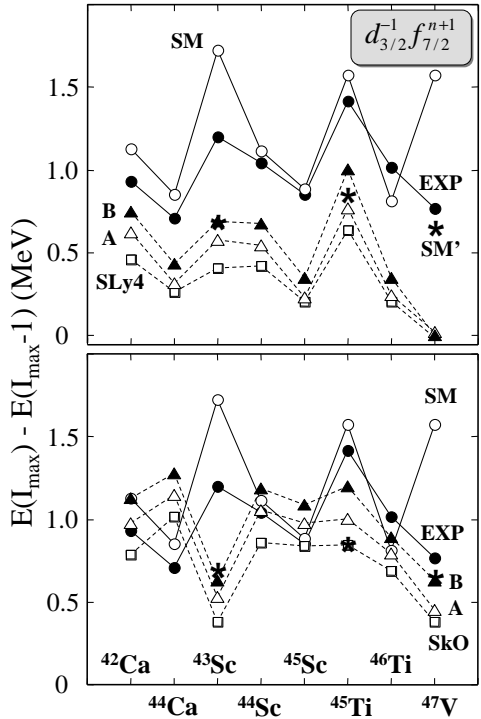}}
\caption{The  energy difference
between $I_{max}$ and $(I_{max}-1)_1$ states in  $d_{3/2}^{-1}f_{7/2}^{n+1}$
configurations in  $A$$\sim$44 nuclei.
Dots and circles mark experimental
data and the SM results, respectively.
The asterisk symbols (SM')
show the SM results for the $(I_{max}-1)_2$ states
in $T$=1/2 Sc, Ti, and V isotopes.
The CHF
results with the SLy4 and SkO functionals are shown in the upper
and lower panel, respectively.
The  lowest cranked solution ($|\bar\pi\rangle$)
is marked by squares while open  (filled) triangles
mark the  results
obtained within the approximate angular
momentum projection A(B) are marked with open (filled)
triangles.}\label{df}
\end{figure}

The SM and CHF results corresponding to the lowest
$(I_{max}-1)_1$ states are shown in
Fig.~\ref{df}. The agreement between SM and experiment is again
excellent, except for  the $T$=1/2 nuclei
$^{43}$Sc and  $^{47}$V where theory overestimates the
 energy difference   between $I_{max}$ and
$I_{max}-1$ intruder states. In order to understand this apparent discrepancy,
Fig.~\ref{df} also shows the SM results for excited $(I_{max}-1)_2$ states
in $T$=1/2 nuclei. It turns out that for $^{47}$V the calculated 33/2$^+_1$
level has a neutron intruder character, while it is  the second
33/2$^+_2$ state in which the proton intruder configuration dominates.
For $^{45}$Ti, the calculated 31/2$^+_{1,2}$ states have a mixed proton-neutron
intruder character with a slight preference for a proton configuration,
 and the same is true for the 25/2$^+_{1,2}$ states in
$^{43}$Sc. Clearly, the energetics of $(I_{max}-1)_1$ and $(I_{max}-1)_2$
states in $T$=1/2 nuclei strongly depends on the mixing between proton
and neutron intruder states, i.e, the  isospin dependence of the
cross-shell $sd\leftrightarrow fp$ interaction.

%

The CHF calculations displayed in Fig.~\ref{df}
were carried out  for the SLy4  (top) and SkO (bottom) Skyrme
parameterizations.
It is interesting to see
that predictions for
 $d_{3/2}^{-1}f_{7/2}^{n+1}$ intruder states strongly depend on the
energy density  functional.
 The mean
empirical value of $\Delta E \equiv
E(I_{max}) - E_1 (I_{max}-1)$, averaged over   all nuclei considered, is
$\overline{\Delta E}_{\textrm{EXP}} \approx 0.990$\,keV.
For the lowest CHF solutions corresponding to $\bar\pi$-type
configurations (marked by squares in the figures), the
calculations yield $\overline{\Delta E}_{\textrm{SLy4}} \approx 0.330$\,keV
and $\overline{\Delta E}_{\textrm{SkO}} \approx 0.730$\,keV.
The calculated corrections are of similar size
for both Skyrme functionals, and in all cases they
improve the agreement with  the data. Within  method A, the CHF
curves are shifted up, on the average, by
$\delta \overline{\Delta E}_{\textrm{SLy4}} \approx 0.090$\,keV
and $\delta \overline{\Delta E}_{\textrm{SkO}} \approx 0.140$\,keV.
The corrections obtained in  variant B are
$\delta \overline{\Delta E}_{\textrm{SLy4}} \approx 0.210$\,keV
and $\delta \overline{\Delta E}_{\textrm{SkO}} \approx 0.270$\,keV.

In variant   B, the average
splitting calculated in CHF-SkO
is fairly close to the experimental value,
$\overline{\Delta E}_{\textrm{SkO}}\approx
\overline{\Delta E}_{\textrm{EXP}}\approx 1$\,MeV.
However, clear discrepancies in the
isotopic and isotonic
dependence are seen.
Applying projection method B to CHF-SLy4  yields
an almost constant offset of
$\overline{\Delta E}_{\textrm{EXP}} - \overline{\Delta E}_{\textrm{SLy4}}
\sim 0.450$\,keV. Indeed, by shifting the theoretical results
up  by  $\sim 0.450$\,keV, one reproduces
surprisingly well the empirical isotopic
and isotonic dependence, including the cases
of $^{43}$Sc and $^{47}$V.
It is important to point out that both methods A and B are free from
adjustable parameters. In both cases we have simplified the
picture by limiting the size of  the configuration
space to three states, and
in both cases we have forced the spurious mode to have zero energy.

\section{Favorite Signature Terminating $I_{max}-2$ states}\label{Im2}
\begin{figure}[htb]
\centerline{\includegraphics[
width=0.45\textwidth,clip]{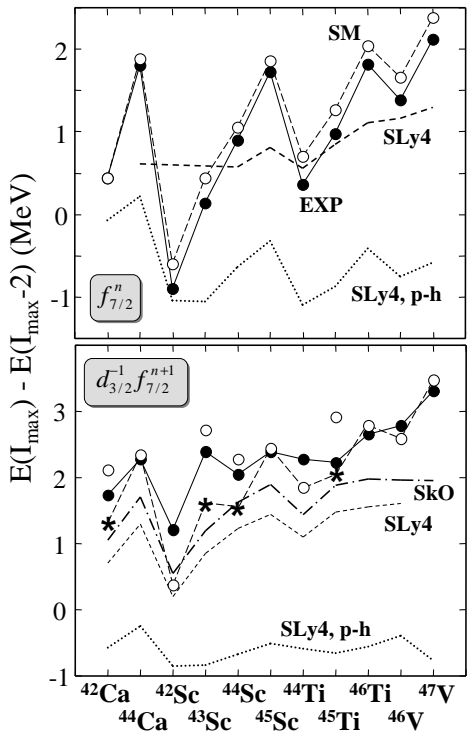}}
\caption{The transition energy
between $I_{max}$ and $(I_{max}-2)_1$ states in  $f_{7/2}^{n}$ (top) and
$d_{3/2}^{-1}f_{7/2}^{n+1}$ (bottom) configurations in  $A$$\sim$44 nuclei.
Dots and circles mark experimental data and the SM results, respectively. The
intruder $(I_{max}-2)_2$ states in  $^{42}$Ca, $^{43,44}$Sc, and $^{45}$Ti are
shown by asterisks. In those nuclei, the $B(E2)$ rates for the
$I_{max}$$\rightarrow$$(I_{max}-2)_2$ transitions are significantly larger
than those for the yrast $I_{max}$$\rightarrow$$(I_{max}-2)_1$ transitions.
The CHF-SLy4 results
are marked by a short-dashed line. The corresponding average  particle-hole
excitation energy is shown by a dotted line. For comparison, the
CHF-SkO results (dash-dotted line) are given in the bottom panel.}\label{DI2}
\end{figure}

In the standard picture of
band termination, the $I_{max}-2$ configurations
contain some quadrupole collectivity. That is,
these states have  small quadrupole deformations that give rise
to a small collective rotational component in the wave function.
One can therefore expect that the energy difference between $I_{max}$ and
$I_{max}-2$ states should mainly depend on  time-even nuclear
multipole  fields associated with the  nuclear deformability.
The  energy of the last quadrupole transition within the terminating
sequence, $\Delta E_2$$\equiv$$E(I_{max})-E(I_{max}-2)$, is shown in Fig.~\ref{DI2}
for both  $f_{7/2}^{n}$ (top) and $d_{3/2}^{-1}f_{7/2}^{n+1}$ (bottom)
structures. The agreement between SM and experiment is excellent.
In  $^{42}$Ca, $^{43,44}$Sc, and
$^{45}$Ti, the $B(E2)$ rates
for the $I_{max}$$\rightarrow$$(I_{max}-2)_2$ transitions are significantly
larger than those for the yrast
$I_{max}$$\rightarrow$$(I_{max}-2)_1$ transitions. In these
cases, the quadrupole strength is fragmented and the band structures cannot be
easily identified.
It is interesting to see that $\Delta E_2$
for the $f_{7/2}^{n}$  configurations is significantly smaller (and sometimes
even negative) as compared to the $d_{3/2}^{-1}f_{7/2}^{n+1}$ values.
This is consistent with the fact that
the quadrupole polarization in the intruder states  is significantly
greater than in the  $f_{7/2}^{n}$ structures.

The mean  p-h $\Delta J_z$=$-$2
excitation energy calculated
in HF-SLy4 lies $\sim$2\,MeV below the data. By performing
self-consistent cranking, one arrives at a slightly collective solution that,
on the average, is  not too far from experiment. For the
weakly collective, normal-parity
structures, the CHF theory poorly reproduces the particle-number dependence. The
situation is significantly improved  in the more collective intruder states where
the mean-field theory (both in CHF-SLy4 and CHF-SkO variants)
performs  much better.
 Here, the energy gain due to deformation
can be as large as $\sim$2\,MeV.
 Of course, a big part of the remaining
discrepancy between CHF results and experiment is due to angular momentum
violation in the $I_{max}-2$ states. Unfortunately,
for those states, the approximate methods of restoring rotational
invariance discussed in the previous sections do not apply because of
many states  involved (i.e., pronounced configuration mixing).
%

\section{Conclusions} \label{conclusion}
In this work we performed a theoretical
analysis of optimal $I_{max}$  states and
non-fully-stretched
$I_{max}-2$ and $I_{max}-1$  states in the $A$$\sim$44 nuclei
from the lower-$fp$ shell.
Overall, the level of agreement between SM results and
experimental data for
$f_{7/2}^n$ seniority isomers
and $d_{3/2}^{-1} f_{7/2}^{n+1}$ intruder
states  is excellent.
We have shown that
CHF solutions for unfavored-signature terminating
states are affected by
dramatic violation of rotational
symmetry even if the shape of nuclei
under consideration is almost spherical. Approximate
methods of restoring rotational invariance in $I_{max}-1$
configurations have been proposed.
The energy
corrections due to angular momentum projection can be
significant, and their inclusion  improve agreement with the data.
Finally, we investigated the weakly collective
$I_{max}-2$  members of terminating structures. The correlation
energy in these states, mainly of  a quadrupole nature,
is fairly large. While for nearly-spherical high-spin
$f_{7/2}^n$  states the CHF method gives only a rough estimate of the
energy splitting between $I_{max}-2$ and $I_{max}$ states, the agreement
is more quantitative for intruder configurations.

\begin{acknowledgments}
This work was supported in part by  the U.S. Department of Energy
under Contract Nos. DE-FG02-96ER40963 (University of Tennessee),
DE-AC05-00OR22725 with UT-Battelle, LLC (Oak Ridge National
Laboratory), DE-FG05-87ER40361 (Joint Institute for Heavy Ion
Research), W-7405-Eng-48 with University of California
(Lawrence Livermore National
Laboratory); by the Polish Committee for
Scientific Research (KBN) under contract No. 1~P03B~059~27; and  by the
Foundation for Polish Science (FNP).
\end{acknowledgments}


\end{document}